\documentclass[sigconf,natbib=true,anonymous=false]{acmart}

\usepackage{booktabs}
\usepackage{amsmath}
\AtBeginDocument{%
  }

\setcopyright{acmlicensed}
\copyrightyear{2026}
\acmYear{2026}
\setcopyright{cc}
\setcctype{by}
\acmConference[CIKM '26]{Proceedings of the 35th ACM International Conference on Information and Knowledge Management}{November 07--11, 2026}{Rome, Italy}
\acmBooktitle{Proceedings of the 35th ACM International Conference on Information and Knowledge Management (CIKM '26), November 07--11, 2026, Rome, Italy}
\acmDOI{10.1145/3799682.3840059}
\acmISBN{979-8-4007-2539-5/2026/11}

\begin{document}

\title{Knowing When to Escalate: Disagreement Patterns for Uncertainty-Aware Content Moderation}

\author{Michał Wawer}
\orcid{0009-0004-2717-1616}
\affiliation{%
  \institution{Laboratory of The New Ethos,\\ Warsaw University of Technology}
  \city{Warsaw}
  \country{Poland}
}
\email{michal.wawer.stud@pw.edu.pl}

\author{Jarosław A. Chudziak}
\orcid{0000-0003-4534-8652}
\affiliation{%
  \institution{Institute of Computer Science,\\ Warsaw University of Technology}
  \city{Warsaw}
  \country{Poland}}
\email{jaroslaw.chudziak@pw.edu.pl}

\renewcommand{\shortauthors}{Michał Wawer and Jarosław A. Chudziak}

\begin{abstract}
    Agentic content-moderation systems increasingly rely on collective judgments from LLM agents, while aggregation methods commonly treat inter-agent disagreement as a reliability problem to be removed through consensus or voting. This framing is inappropriate for value-laden tasks, where disagreement may reflect irreducible value pluralism - a distinction standard uncertainty decompositions cannot recover, since they operate on output distributions and discard the reasoning behind each judgment. We propose Reasoning Trace Divergence Analysis (RTDA), which embeds model-generated justifications from perspective-differentiated agents and classifies each case by justification similarity and conclusion agreement. The resulting four-state taxonomy isolates convergent disagreement, where agents reasoning similarly but concluding differently, as a candidate signature of value-laden conflict. On 1,000 items from the Measuring Hate Speech corpus, this structural signal predicts human annotator disagreement better than magnitude-only and output-only baselines.  We contribute a design principle: condition human escalation on the structure of disagreement, not its magnitude alone, with RTDA serving as a post-processing layer for systems that already generate reasoning traces.
\end{abstract}

\ccsdesc[500]{Computing methodologies~Multi-agent systems}

\keywords{reasoning traces, content moderation, collective intelligence, human-in-the-loop, multi-agent systems, large language models}

\maketitle

\section{Introduction}

Content moderation demands decisions at scale across content that spans a spectrum from unambiguous to genuinely contested \citep{gorwa2020algorithmic}. A comment referencing cultural practices may appear offensive to moderators unfamiliar with the context but appropriate to those within the community. Such disagreement can reflect legitimate value differences, but it can also arise from missing context, unclear policies, item quality, or annotator noise \citep{gillespie2018custodians,aroyo2023dices}.

Multi-agent architectures have emerged as a promising approach, leveraging collective reasoning among agents to improve decision quality \citep{du2024improving, wang2025mixtureofagents,Wawer_Gajewska_Budzynska_Chudziak_2026}. However, current implementations treat inter-agent disagreement primarily as a problem to overcome through consensus protocols \citep{zheng2025rethinkingreliabilitymultiagentsystem} or weighted voting \citep{jo2025byzantine}. This framing implicitly assumes the existence of ground truth that disagreeing agents are failing to reach. Building on our preliminary workshop study \citep{wawer2026when}, the present work adds disjoint calibration and test sets, six baselines, statistical analysis and cross-model and agent-composition ablations. 

We examine whether this assumption is too restrictive for value-laden classification. In particular, we test whether agents that produce semantically similar justifications yet reach different conclusions identify content on which human annotators also tend to disagree. This association can provide a useful escalation signal even when its underlying cause cannot be identified from model outputs alone.

\begin{figure}[t]
    \centering
    \includegraphics[width=0.95\linewidth]{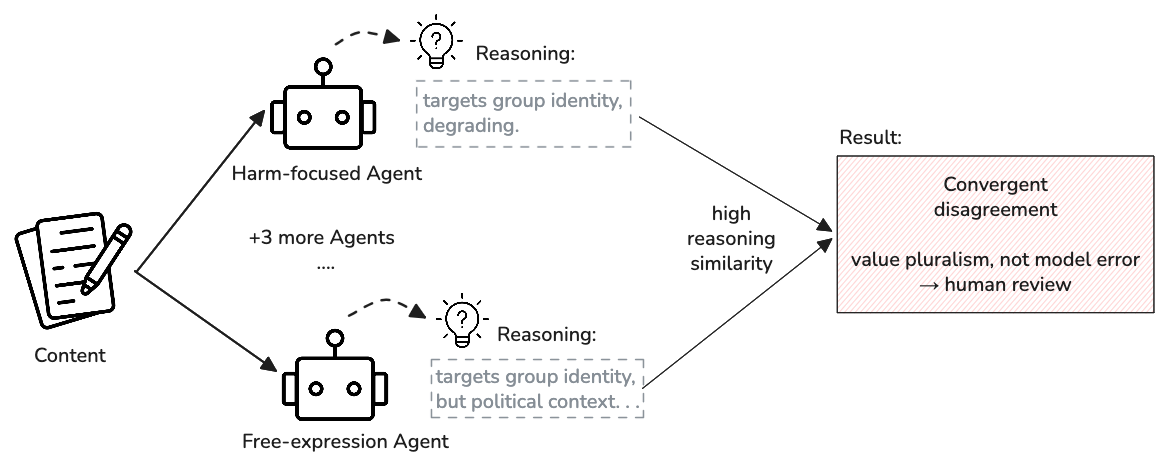}
    \caption{Convergent disagreement: when agents reason similarly but conclude differently. We study this pattern as a candidate signal for human review.}
    \label{fig:teaser}
\end{figure}

Drawing on uncertainty quantification and data perspectivism, we test whether model-generated justifications contain routing information that output distributions discard. Our contribution is a escalation framework that operationalizes this four-state taxonomy using justification similarity and conclusion agreement. It identifies cases where agents verbalize similar considerations yet reach opposing conclusions as candidates for human review. Figure \ref{fig:teaser} illustrates the convergent-disagreement pattern that motivates our structural routing signal. On the evaluated sample, taxonomy-based routing outperforms magnitude-only and output-only baselines, while cross-model ablations test whether the pattern is tied to one model. The taxonomy requires no task-specific training data and can be applied as a post-processing layer when a system already collects structured reasoning traces.

\section{Related Work}\label{sec:related}

Multi-agent debate can improve factuality and reasoning \citep{du2024improving}, but recent work reveals systematic limitations. When agents share training data, debate converges toward shared misconceptions \citep{estornell2024multi} and heterogeneous debate can degrade performance through sycophantic agreement that worsens with longer exchanges \citep{wynn2025talk}. 
Recent domain-specific systems likewise use structured agent differentiation \citep{szczepanik2025collaborativellmagentsc4} and identity-grounded ethical tutoring \citep{maslowski2026heterogeneousdebateengineidentitygrounded} illustrate the broader use of role-conditioned panels in complex reasoning tasks.
LLM groups reproduce human collective-reasoning failures, reaching premature consensus within two rounds \citep{hiddenbench2026}. These findings suggest the absence of disagreement may itself be a failure mode, motivating analysis of disagreement structure rather than its suppression.

The data perspectivism movement challenges the assumption that annotation disagreement reflects error, distinguishing the mere collection of disaggregated labels from their preservation throughout the pipeline \citep{basile2023perspectivist} and showing that disagreement carries different meaning under descriptive versus prescriptive annotation \citep{rottger2022two}. Surveys of learning-from-disagreement methods find that preserving disagreement generally matches or improves on gold-standard training \citep{uma2021learning,leonardelli2023lewidi} and annotator composition alone can alter a non-trivial fraction of outcomes \citep{gordon2022jury}. We extend this tradition to multi-agent systems, treating agent disagreement as an analog of annotator disagreement.
The standard information-theoretic decomposition splits predictive uncertainty into aleatoric and epistemic components \citep{hullermeier2021aleatoric}, but this binary is increasingly contested: it violates intuitive axioms \citep{wimmer2023quantifying}, the two estimates correlate strongly in practice \citep{mucsanyi2024benchmarking}, and richer taxonomies have been proposed \citep{bickford2025rethinking}. 

Content moderation involves genuine normative tensions rather than factual uncertainty alone \citep{gillespie2018custodians,gorwa2020algorithmic}. Safety judgments are socio-culturally situated with substantial disagreement between expert labels and diverse crowd judgments \citep{aroyo2023dices}, standard RLHF narrows this distributional pluralism \citep{sorensen2024roadmap} while divergent community-model outputs can serve as a direct pluralism signal \citep{feng2024modular}. We operationalize these insights more narrowly: disagreement structure is evaluated as a predictor of when pluralistic human review may be useful, without assuming that agent disagreement and human value pluralism are the same construct.

\section{Theoretical Framework}\label{sec:framework}

We frame multi-agent moderation as judgment under uncertainty and motivate a structural routing hypothesis. For the analysis that follows, we operationalize reasoning traces as the textual justifications generated by agents alongside their moderation decisions. 

\subsection{Uncertainty in Multi-Agent Judgment}

We frame multi-agent content moderation as a decision problem under uncertainty. Let $\mathcal{X}$ denote the space of content items and $\mathcal{Y} = \{0, 1\}$ the binary moderation decision. A panel of $N$ agents $\{a_1, \ldots, a_N\}$ each produce a decision $\hat{y}_i \in \mathcal{Y}$ and a reasoning trace $r_i \in \mathcal{R}$ for each content item $x \in \mathcal{X}$.

Standard ensemble uncertainty decomposition \citep{hullermeier2021aleatoric} splits total predictive uncertainty $\mathcal{H}[Y \mid x]$ into:
\begin{equation}\label{eq:decomp}
    \underbrace{\mathcal{H}[\bar{p}(y \mid x)]}_{\text{total}} = \underbrace{\mathbb{E}_{\theta}[\mathcal{H}[p_\theta(y \mid x)]]}_{\text{aleatoric (agent entropy)}} + \underbrace{\mathcal{I}[Y; \Theta \mid x]}_{\text{epistemic (agent disagreement)}}
\end{equation}
where $\bar{p}$ is the ensemble average, $\theta$ indexes individual agents, and $\mathcal{I}$ denotes mutual information. This decomposition treats all inter-agent disagreement as epistemic uncertainty---information that could be resolved with better models.

For value-laden tasks, this output-only view may be incomplete. Agent disagreement can reflect competing value priorities, resolvable knowledge gaps, missing context, prompt sensitivity, or stochastic variation. Equation~\ref{eq:decomp} cannot distinguish among these possibilities because it operates only on output distributions. We therefore test whether the structure of observable justifications is useful for routing, rather than claiming to recover the latent cause of disagreement.

For agent panel $\{a_1, \ldots, a_N\}$ with decisions $\hat{\mathbf{y}} = (\hat{y}_1, \ldots, \hat{y}_N)$ and reasoning traces $\mathbf{r} = (r_1, \ldots, r_N)$ on content $x$, we define:
\begin{itemize}
    \item \textbf{Conclusion agreement}: let
    \begin{equation}
        q(x)=\max_{y\in\mathcal{Y}}\frac{1}{N}\sum_{i=1}^{N}\mathbb{1}[\hat{y}_i=y],
    \end{equation}
    then $\mathcal{A}_{\tau_a}(x)=\mathbb{1}[q(x)\geq\tau_a]$.
    \item \textbf{Reasoning similarity}: 
    \begin{equation}
        \mathcal{S}(x)
        =
        \frac{2}{N(N-1)}
        \sum_{1 \leq i < j \leq N}
        \operatorname{sim}(r_i,r_j),
        \label{eq:reasoning-similarity}
    \end{equation}
    where $\operatorname{sim}(r_i,r_j)$ denotes the cosine similarity between
    the embedded reasoning traces of agents $i$ and $j$.
\end{itemize}

\subsection{Four-state (dis)agreement taxonomy}

We classify each content item into categories based on the joint distribution of $\mathcal{A}_{\tau_a}(x)$ and $\mathcal{S}(x)$ relative to thresholds $\tau_a$ and $\tau_s$:
\begin{align}
    \textsc{CA}(x) &: \mathcal{S}(x) \geq \tau_s \;\wedge\; \mathcal{A}_{\tau_a}(x) = 1 \quad \text{(Convergent Agreement)} \nonumber \\
    \textsc{DA}(x) &: \mathcal{S}(x) < \tau_s \;\wedge\; \mathcal{A}_{\tau_a}(x) = 1 \quad \text{(Divergent Agreement)} \nonumber \\
    \textsc{CD}(x) &: \mathcal{S}(x) \geq \tau_s \;\wedge\; \mathcal{A}_{\tau_a}(x) = 0 \quad \text{(Convergent Disagreement)} \nonumber \\
    \textsc{DD}(x) &: \mathcal{S}(x) < \tau_s \;\wedge\; \mathcal{A}_{\tau_a}(x) = 0 \quad \text{(Divergent Disagreement)} \nonumber
\end{align}

Each category maps to a distinct uncertainty interpretation:

\textbf{Convergent Agreement} (high similarity, same conclusion): Agents reason about the content in compatible ways and reach the same judgment. Standard ensemble uncertainty is low. These cases are candidates for automated decisions.

\textbf{Divergent Agreement} (low similarity, same conclusion): Agents reason differently but converge on the same judgment through independent paths. This pattern is consistent with conclusion stability across prompted perspectives, analogous to ``diverse ensembles'' in the wisdom-of-crowds literature, but does not by itself establish robustness to other models or policies \citep{page2008difference,schoenegger2024wisdom}.

\textbf{Convergent Disagreement} (high similarity, different conclusions): Agents share a compatible understanding of the content but reach different judgments. This is the signature of irreducible value pluralism. The agents agree on what the content says but disagree on how it should be moderated. In the uncertainty decomposition framework, this corresponds to uncertainty that cannot be resolved by improving the model, only by specifying which values to prioritize. This is exactly where human judgment is needed.

\textbf{Divergent Disagreement} (low similarity, different conclusions): Agents interpret the content differently and reach different judgments. This pattern may reflect epistemic uncertainty (addressable through better context or models), content ambiguity, or noise from inconsistent interpretation.

\subsection{Scope of Agent Diversity}

Our agents are homogeneous at the model-backend level and differentiated through perspective prompts. All agents within a panel share the same learned knowledge and architecture, while their system prompts emphasize different moderation considerations. This design provides controlled perspective variation but may retain shared blind spots and conformity tendencies \citep{zhu2025conformity,chuang2025debate}.

We therefore interpret disagreement operationally, as evidence that a decision is sensitive to the prompted perspective, rather than as proof of human-like value pluralism. Such sensitivity may nevertheless be useful for escalation because it identifies cases whose outcomes vary across relevant moderation viewpoints. The cross-model ablation tests whether this routing pattern recurs across separate models.

\section{Approach and Methods}
\label{sec:method}

This section presents the taxonomy as a pipeline: perspective differentiated deliberation, reasoning-trace embedding and four-state classification.

\subsection{Perspective-Differentiated Deliberation}
\label{sec:agents}

We instantiate five agents from the same base model, differentiated by system prompts encoding distinct moderator perspectives: \textbf{harm-focused} (protection from psychological harm and harassment), \textbf{context-sensitive} (cultural and situational interpretation), \textbf{community-norms} (platform-specific behavioral standards), \textbf{free-expression} (speech protection against restriction), and \textbf{legal-framework} (jurisdictional standards and precedent).

This design creates controlled, prompt-level perspective variation while holding the backend constant. We therefore do not call the agents model-heterogeneous. Agents independently generate moderation decisions and structured justifications through chain-of-thought prompting \citep{wei2022chain}, articulating: (1) content interpretation, (2) relevant considerations, (3) value trade-offs, and (4) final judgment with confidence.

\subsection{Reasoning Trace Embedding}

Following \citet{tajik2026disagreement}, who demonstrated that cosine similarity of reasoning traces robustly differentiates consensus from disagreement in multi-agent coding tasks, we embed traces into a shared semantic space for quantitative comparison.

For each agent $i$ and content item $c$, let $r_i^c$ denote the full reasoning trace. We compute sentence embeddings using a pre-trained transformer model (all-mpnet-base-v2), yielding vector representations $\mathbf{e}_i^c \in \mathbb{R}^d$. Pairwise cosine similarity is:
\begin{equation}
    s_{ij}^c = \frac{\mathbf{e}_i^c \cdot \mathbf{e}_j^c}{\|\mathbf{e}_i^c\| \, \|\mathbf{e}_j^c\|}
\end{equation}

From pairwise similarities, we derive aggregate metrics for each content item:
\begin{align}
    \bar{s}^c &= \frac{2}{N(N{-}1)} \sum_{i<j} s_{ij}^c \quad \text{(mean similarity)} \\
    s_{\min}^c &= \min_{i<j} s_{ij}^c \quad \text{(minimum similarity)} \\
    \sigma_s^c &= \text{Var}(\{s_{ij}^c\}_{i<j}) \quad \text{(similarity variance)}
\end{align}

Low mean similarity indicates dissimilar generated justifications. High mean similarity with conclusion disagreement indicates that agents verbalized overlapping considerations but applied them differently. These interpretations concern text-level similarity, not direct access to latent reasoning.

\subsection{Taxonomy Classification}

We set $\mathcal{A}_{\tau_a}(x)=1$ when the majority agreement rate $q(x)$ reaches $\tau_a$, and set $\mathcal{S}(x)=\bar{s}^{c}$. We fix $\tau_a=0.80$ a priori, corresponding to agreement by at least four of five agents. The similarity threshold $\tau_s=0.72$ was selected using the
200-item calibration set according to the procedure described in Section~\ref{sec:setup}. Both thresholds were fixed before evaluation on the test set.

We separate the cost of producing the agent panel from the incremental cost of RTDA. For a panel of $N$ agents, the pipeline requires $N$ LLM generations, $N$ trace embeddings, and $\binom{N}{2}$ cosine comparisons per item. With $N=5$, this corresponds to five generations, five embeddings, and ten comparisons, or 5,000 generations in the 1,000-item headline experiment. Once the justifications are available, RTDA adds only the embedding and comparison stages. For comparison, B4 and B5 each require ten sampled generations per item.

\section{Experiments}
\label{sec:experiments}
 
We test the narrower claim that the structure of inter-agent disagreement is a predictor of human annotator disagreement and a useful escalation signal. We conduct: (i) a comparison against magnitude-only, output-only and supervised baselines with a threshold sweep; and (ii) ablations over base models and agent compositions.
 
\subsection{Setup and Evaluation Criteria}
\label{sec:setup}
 
We evaluate on the Measuring Hate Speech corpus~\cite{kennedy2020constructing,sachdeva2022measuring}, which preserves individual annotator labels rather than collapsing them into a single gold label. We sample a supervised-training set ($n_{\mathrm{train}}=200$), a calibration set ($n_{\mathrm{cal}}=200$), and a held-out test set ($n_{\mathrm{test}}=1{,}000$). Per-item human disagreement $\bar{d}_h(c)\in[0,1]$ is computed as the normalized standard deviation of annotator ratings. Items in its top quartile are designated high annotator-disagreement cases for routing. This is an operational proxy for review priority, not a direct label of value pluralism or item ambiguity.
 
For each case $c$, $k=5$ perspective-differentiated LLM agents (described in  Section~\ref{sec:agents}) generate a structured output $\langle r_i, d_i, \gamma_i\rangle$ where $r_i$ is the reasoning trace, $d_i\in\{\textsc{Keep},\textsc{Remove}\}$, and $\gamma_i\in[0,1]$ is verbalized confidence. Reasoning traces are embedded with all-mpnet-base-v2 and mean pairwise cosine similarity $\overline{sim}(c)$ is computed across the $\binom{5}{2}=10$ pairs. The four-state taxonomy $\Phi$ assigns each case to one of $\{CA,DA,DD,CD\}$ via thresholds $(\tau_s,\tau_a)$ as defined in Section~\ref{sec:method}.

For the headline experiment we use GPT-5.4-Nano. For the model-family ablation we repeat the pipeline with Claude~4.5 Sonnet and MiniMax M3, holding the five prompts and decoding configuration fixed. All code, prompts, configurations, and evaluation scripts are publicly available \footnote{\url{https://github.com/michal-wawer/agentic-reasoning-traces-analysis-in-content-moderation}}. Operating thresholds and F1 decision points were selected only on held-out calibration data and then fixed for evaluation. 

The calibration set was used to evaluate $\tau_s\in\{0.60-0.85\}$. We selected $\tau_s=0.72$ because it maximized F1 for identifying high annotator-disagreement items on this set. Method-specific operating thresholds for B1-B5 and the B6 decision threshold were selected using the same criterion on the calibration set. After calibration, all thresholds were fixed before evaluation. 

We report two tasks: (i) Predictive validity: Pearson and Spearman correlation between an agent-side disagreement signal and $\bar{d}_h(c)$; (ii) Routing: binary classification of high annotator-disagreement items, reported as AUROC and F1 at a calibration-selected operating point. We use two-sided tests and Fisher's $z$ transformation for 95\% confidence intervals on Pearson correlations. These intervals characterize item-sampling uncertainty, stochastic variation across repeated LLM runs was not measured.

We compare against six baselines. 
(B1) Mean reasoning-trace divergence $1-\overline{sim}(c)$. 
(B2) Decision-vote entropy across the five agents.
(B3) Single-agent verbalized confidence~\cite{lin2022teachingmodelsexpressuncertainty}. 
(B4) Self-consistency confidence~\cite{wang2023selfconsistencyimproveschainthought}: agreement rate across $K=10$ sampled completions of a single agent at $T{=}0.7$. 
(B5) Semantic uncertainty~\cite{kuhn2023semanticuncertaintylinguisticinvariances}: predictive entropy over meaning-equivalence clusters from $K=10$ samples.
(B6) A supervised gradient-boosted classifier (LightGBM) trained on the 200-item supervised-training set using $\overline{sim}(c)$, majority agreement, vote entropy, mean verbalized confidence, and pairwise-similarity statistics. B6 alone uses human-disagreement labels during training; all other methods are unsupervised.

\subsection{Structure vs.\ Magnitude}
\label{sec:headline}
 
Table~\ref{tab:headline} reports correlations and routing performance on the Measuring Hate Speech sample.
 
\begin{table}[h]
\centering
\small
\caption{Predictive validity and routing performance. $r$, $\rho$: Pearson and Spearman correlation with human annotator disagreement, $^{\dagger}$ denotes supervised baseline.}
\label{tab:headline}
\begin{tabular}{lcccc}
\toprule
Method & $r$ & $\rho$ & AUROC & F1 \\
\midrule
B1 Divergence-only         & 0.38 & 0.41 & 0.68 & 0.51 \\
B2 Vote entropy            & 0.34 & 0.36 & 0.66 & 0.50 \\
B3 Verbalized conf.        & 0.26 & 0.29 & 0.61 & 0.47 \\
B4 Self-consistency        & 0.31 & 0.33 & 0.64 & 0.49 \\
B5 Semantic uncertainty    & 0.35 & 0.37 & 0.67 & 0.51 \\
B6 Supervised$^{\dagger}$  & 0.43 & 0.45 & 0.73 & 0.56 \\
\midrule
\textbf{Structural ($\Phi$)} &
  \textbf{0.41} & \textbf{0.44} & \textbf{0.72} & \textbf{0.59} \\
\bottomrule
\end{tabular}
\end{table}
 
The structural signal $\Phi$ outperforms all magnitude-only and single-model uncertainty baselines on every metric, and is competitive with the supervised classifier (B6) without requiring labeled training data. Its Pearson association with human disagreement is $r{=}0.41$ (95\% CI $[0.36,0.46]$). Within the convergent-disagreement subset (Table~\ref{tab:cd-subset}), the association is stronger ($r{=}0.52$, 95\% CI $[0.38,0.64]$; $\rho{=}0.55$). These comparisons demonstrate association and routing utility, not the cause of human disagreement.
 
\begin{table}[h]
\centering
\caption{Per-state distribution and Pearson correlation with human annotator disagreement on Measuring Hate Speech ($n=1{,}000$); brackets give Fisher-$z$ 95\% confidence intervals.}
\label{tab:cd-subset}
\begin{tabular}{lccc}
\toprule
State & $n$ & Mean $\bar{d}_h$ & $r$ [95\% CI] \\
\midrule
$DA(c)$ & 197 & 0.34 & 0.19 $[0.05,0.32]$ \\
$CA(c)$ &  41 & 0.59 & 0.27 $[-0.04,0.53]$ \\
$DD(c)$ & 635 & 0.73 & 0.38 $[0.31,0.44]$ \\
$CD(c)$ & 127 & 0.81 & \textbf{0.52} $[0.38,0.64]$ \\
\bottomrule
\end{tabular}
\end{table}

The corresponding two-sided $p$-values are $0.008$ (DA), $0.088$ (CA), and $<.001$ (DD and CD). The wide interval for CA follows from its small subset size ($n=41$), reinforcing that per-state estimates should be interpreted cautiously.
 
\begin{figure}[b]
    \centering
    \includegraphics[width=0.95\linewidth]{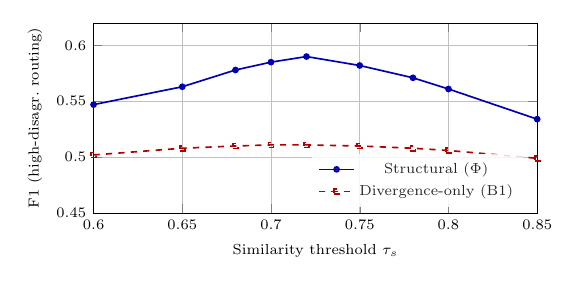}
    \caption{F1 over the similarity threshold $\tau_s$ at fixed $\tau_a=0.80$. The structural advantage persists across an interval rather than at a single operating point.}
    \label{fig:threshold-sweep}
\end{figure}

The headline results use $\tau_s=0.72$, selected by maximizing routing F1 on the calibration set, and $\tau_a=0.80$, fixed a priori to require agreement by at least four of five agents.
As a post-hoc sensitivity analysis, Figure~\ref{fig:threshold-sweep} reports test-set F1 across the predefined candidate values of $\tau_s$ while holding $\tau_a=0.80$ fixed. This test-set sweep was not used to alter the headline operating point. Across $\tau_s\in[0.65,0.80]$, the structural method remains at least five F1 points above divergence-only.

\subsection{Ablations} 
\label{sec:ablations}
 
We isolate two sources of variability: the base LLM and the agent composition.
 
\paragraph{Base-model ablation.}
Table~\ref{tab:model-ablation} repeats the headline pipeline with three separate models. The qualitative ordering ``structural outperforming magnitude-only`` is preserved across all three, and absolute F1 changes by at most $3$ points. The within-state ordering (CD highest, DA lowest) also repeats. This supports cross-backend robustness of the observed association, but it does not demonstrate heterogeneity among agents within a panel.
 
\begin{table}[h]
\centering
\small
\caption{Base-model ablation: routing F1 on Measuring Hate Speech, fixed agent prompts.}
\label{tab:model-ablation}
\begin{tabular}{lcccc}
\toprule
Base model & $\Phi$ & B1 div.\ & $\Delta$ F1 & $r$ on CD \\
\midrule
GPT-5.4-Nano    & 0.59 & 0.51 & +0.08 & 0.52 \\
Claude 4.5 Son. & 0.57 & 0.50 & +0.07 & 0.49 \\
MiniMax M3.     & 0.58 & 0.49 & +0.09 & 0.51 \\
\bottomrule
\end{tabular}
\end{table}
 
\paragraph{Agent-composition ablation.}
We vary the agent set along two axes (Table~\ref{tab:composition-ablation}): (i) the number of agents $k\in\{3,5,7\}$; and (ii) prompt profiles, by dropping or replacing a perspective and by removing differentiation entirely. With five identical prompts and only decoding seeds varied, $\Delta$F1 falls to $-0.01$. This ablation supports the narrower conclusion that the gain depends on prompt-level perspective differentiation, not merely on sampling multiple outputs.
 
\begin{table}[h]
\centering
\small
\caption{Agent-composition ablation (GPT-5.4-Nano). $\Delta$F1 reported against B1 divergence-only at matching $k$.}
\label{tab:composition-ablation}
\begin{tabular}{lcc}
\toprule
Configuration & F1 ($\Phi$) & $\Delta$ F1 vs.\ B1 \\
\midrule
Full 5 perspectives            & 0.59 & +0.08 \\
$k=3$ (drop legal, community)  & 0.55 & +0.06 \\
$k=7$ (add procedural, victim) & 0.60 & +0.08 \\
Random perspective swap        & 0.56 & +0.05 \\
Drop harm-focused              & 0.53 & +0.04 \\
Drop free-expression           & 0.54 & +0.05 \\
No differentiation             & 0.50 & -0.01 \\
\bottomrule
\end{tabular}
\end{table}

\subsection{Qualitative Examination}

To complement the aggregate metrics, we examine one CD item as an illustrative case: {``These people have completely lost touch with reality and are destroying everything we built.''} The agents identified overlapping features: a group reference, degrading language, and political context - but weighted them differently. The harm-focused and community-norms agents recommended removal, whereas the free-expression and legal-framework agents recommended keeping the item; normalized human disagreement was $\bar d_h=0.82$. Thus, the conclusion divergence did not arise from wholly unrelated generated justifications.

The case nevertheless supports more than one interpretation. The different decisions are consistent with competing priorities between protection from group-directed harm and preservation of political expression. At the same time, the underspecified referent and missing conversational context could have contributed to the disagreement, while generated justifications may rationalize decisions post hoc. RTDA cannot determine which explanation is correct. Its role is therefore to surface a complex case for human review while preserving information about how agents framed the decision, rather than diagnosing the disagreement as irreducible value conflict.

\section{Limitations}

Our evidence is intentionally limited to predictive validity. First, generated justifications may be post-hoc, prompt-sensitive rationalizations. Semantic similarity measures their surface-level content and not the models latent decision processes. Second, high annotator disagreement is an imperfect target. It can reflect competing values, but also unclear items, missing conversational context, inconsistent labels, or annotator noise. RTDA therefore cannot determine why humans disagree.

Third, the evaluation uses a stratified sample of 1,000 English items from one 2022 hate-speech corpus. The sample does not establish robustness across moderation policies, languages, cultures, or domains. Because the corpus is public and predates the evaluated models, pretraining exposure cannot be excluded. The results should not be interpreted as a clean test of uncontaminated zero-shot generalization. Fourth, agents within each panel share a backend and differ through prompts, so their diversity is controlled but not architectural or knowledge-base heterogeneity. Finally, routing quality is evaluated against an annotator-disagreement proxy rather than through a study with professional moderators, and run-to-run stochastic uncertainty and end-to-end operational cost require further measurement. These limitations motivate cross-corpus replication, repeated-run analysis and an actual escalation study before deployment claims are warranted.

\section{Conclusion and Future Work} 

The central contribution of this work is a four-state taxonomy over justification similarity and conclusion agreement, operationalized through RTDA as a post-processing method for multi-agent moderation. On the test set, this structural representation identifies high annotator-disagreement items more effectively than magnitude-only and output-only uncertainty baselines. The composition ablation indicates that this advantage depends on prompt-level perspective differentiation rather than merely sampling multiple outputs, while the same qualitative ordering across separate LLMs suggests that the result is not specific to one backend. By distinguishing how agents disagree rather than only how much, RTDA turns disagreement into an actionable signal for allocating limited human review. This supports our practical design principle that escalation should be conditioned on the structure of inter-agent disagreement, not its magnitude alone.

Future work should first test external validity across additional moderation corpora, languages, cultural settings and platform policies, preferably including data unlikely to have appeared in model pretraining. A human-in-the-loop study should then compare RTDA-assisted escalation with confidence and vote-based routing, measuring decision quality, consistency, reviewer workload, and review time. Repeated stochastic runs and panels containing genuinely different model backends would help establish the stability of the taxonomy, while alternative rationale formats, embedding models, and similarity measures would test whether the signal depends on the current implementation. Finally, learning platform-specific thresholds or a complete state-to-action policy from observed escalation outcomes could adapt RTDA to different review budgets and risk tolerances.


\newpage
\section*{GenAI Usage Disclosure}

We used a generative AI assistant to suggest phrasing, improve
clarity and identify weaknesses in draft text. All research contributions, methods, results, and interpretations are our own.

\bibliographystyle{ACM-Reference-Format}
\bibliography{references}


\end{document}